\documentclass[aps,showpacs,twocolumn]{revtex4}
\usepackage{amsfonts}
\usepackage{dsfont}
\usepackage{amssymb}
\usepackage{amsmath}
\usepackage{graphicx}
\usepackage{epsfig}

\begin{document}

\title{Robust Multiple-Range Coherent Quantum State Transfer}
\author{Bing~Chen$^{1,2}$}
\email{chenbingphys@163.com}
\author{Yan-Dong~Peng$^{1}$}
\author{Yong~Li$^{2,3}$}
\author{Xiao-Feng~Qian$^{4}$}
\email{xfqian@pas.rochester.edu}
\affiliation{$^{1}$College of Electronics, Communication \& Physics, Shandong University
of Science and Technology, Qingdao 266510, China}
\affiliation{$^{2}$Beijing Computational Science Research Center, Beijing 100094, China}
\affiliation{$^{3}$Synergetic Innovation Center of Quantum Information and Quantum
Physics,\\
University of Science and Technology of China, Hefei, Anhui 230026, China}
\affiliation{$^{4}$Center for Coherence and Quantum Optics, The Institute of Optics,
University of Rochester, Rochester, NY 14627, USA}
\date{\today }

\begin{abstract}
We propose a multiple-range quantum communication channel to realize
coherent two-way quantum state transport with high fidelity. In our scheme,
an information carrier (a qubit) and its remote partner are both
adiabatically coupled to the same data bus, i.e., an \emph{N}-site
tight-binding chain that has a single defect at the center. At the weak
interaction regime, our system is effectively equivalent to a three level
system of which a coherent superposition of the two carrier states
constitutes a dark state. The adiabatic coupling allows a well controllable
information exchange timing via the dark state between the two carriers.
Numerical results show that our scheme is robust and efficient under
practically inevitable perturbative defects of the data bus as well as environmental dephasing noise.
\end{abstract}

\maketitle

\section{Introduction}

Quantum state transfer (QST) in many-body solid state physical systems plays
a central role in the realization of various localized quantum computation
and quantum communication proposals \cite{NC-Preskill, bose07}. A practical
high quality quantum state transfer scheme needs to possess several
desirable features: i) high fidelity (to preserve the transferred message),
ii) robustness (to tolerate inevitable practical errors, imperfections, and decoherence),
iii) efficiency (to achieve optimal results with minimal implementations),
and iv) flexibility (to serve for multiple tasks). The investigation of
accomplishing high fidelity quantum state transfer in electronic and spin
systems has recently drawn tremendous attention (see for example \cite{BOSE03, MC04, Subrahmanyam04, Bose05, Li-etal05, Shi-etal05,
Paternostro-etal05, Qian-etal05, Kay07, Franco-etal08, Gualdi-etal08,
Chen-etal14, Lorenzo-etal15, Mitra15} and an overview \cite{bose07}). Many
of these schemes are based on the natural dynamical evolution of permanent
coupled chain of quantum systems, and require no control during the QST.
However, such schemes rely on precise manufacture of the system interaction
parameters as well as accurate timing of information processing, and may not
be robust against experimental imperfection settings, such as small
variations of the system Hamiltonian, environmental noise, etc.

Recently, adiabatic passage has been paid much attention for quantum
information transfer in various physical systems \cite{RMP98, ADG04, KE04,
NV06, SM10, WM07, SL1, SL2, GD, KE06, TO09, TM11, TO07, ADG14, JF05, CJ12,
BEC1, BEC2, BEC3, JA14, ADG06, LMJ09, LCL06, SL14, LMJ10, RM14, DP10, UF15,
EF15}. One typical way is called coherent QST which involves a quantum
system that has an instantaneous eigenstate that is a superposition of a
message state and its corresponding target state. Stimulated Raman adiabatic
passage (STIRAP) \cite{RMP98} is such an example in a three-level atomic
system. In this technique, the dark state which is a coherent superposition
of message and target states plays a central role in the process of
information transfer. In STIRAP evolution, the message and target states are
coupled to a same intermediate state by a pump pulse and a Stokes pulse
respectively. If the two pulses are applied counter-intuitively, i.e., the
Stokes pulse is applied before the pump, then the dark state is associated
initially with the message state and eventually with the target state. Such
a process effectively transports the information of the message state to the
desired target state. There are a couple of advantages of the adiabatic
passage scheme: it is robust against small errors and imperfections of
settings, and the QST timing can be controlled freely and precisely.

In this paper we extend the STIRAP protocol to an \emph{N}-site
tight-binding model and show that it is suitable for high fidelity robust
multiple-range QST and quantum information swapping (see schematic
illustration in Fig. 1). The tight-binding quantum dot (QD) array with a
single diagonal defect serves as the adiabatic pathway for two-way
electronic transport. Two external QDs A, B represent information sender and
receiver or vice versa are allowed to be flexibly side coupled to the array
(i.e., QDs A B can couple to different sites of the QD array as required by
particular tasks). In this scheme, the coupling parameters between the
external QDs and the corresponding sites on the array are made time
dependent, which are controlled by the sender and receiver, respectively. We
find that the ground state of the array is a bound state (or localized
state) due to the existence of the defect. This allows us to show
analytically that our system is an effective three-level system when the
coupling between QD A (B) and the QD array are weak. As a consequence it is
demonstrated that high fidelity two-way QST can be realized at various
different distances between the sender and receiver. Numerical results are
also performed to illustrate that our scheme is robust against dephasing and small
variations of the QD couplings and imperfections.

The paper is organized as follows. In Sec. II, the driven model is described
and the tight-binding QD array system is analyzed in detail. In Sec. III,
the multiple-range adiabatic transport scheme and its effective Hamiltonian are discussed
analytically. A practically imperfect model, as well as an open dephasing model are considered numerically. The last section summarizes the paper.

\begin{figure}[tbp]
\center
\includegraphics[bb=98 345 382 402, width=8 cm, clip]{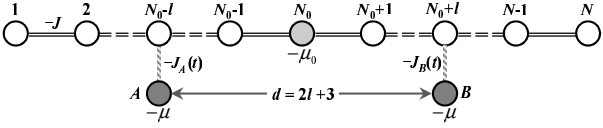}
\caption{Schematic illustrations of multiple-range adiabatic quantum state
transfer from A to B. The tight-binding array with single defect is acting
as a quantum data bus, in which the coupling strengthes $-J$ are
time-independent and the defect QD is supplied with energy $-\protect\mu_{0}$%
. The sender (QD \emph{A}) and the receiver (QD \emph{B}) supplied with
on-site energy,$-\protect\mu$ are coupled to two sites of the array on
opposite sides with respect to the defect site. The sender controls $%
-J_{A}(t)$ and the receiver controls $-J_{B}(t)$. The transfer distance in
terms of the number of sites is given as $d=2l+3$.}
\label{fig1}
\end{figure}

\section{Driven Model}

We start here with the structure of our proposal illustrated in figure 1:
the channel connecting the two side QDs \textit{A} and \textit{B} is a
one-dimensional tight-binding array with uniform and always-on exchange
interactions and with one diagonal defect at $N_{0}$th site. The coupling
between the QDs A, B and their corresponding channel sites are made time
dependent. The total Hamiltonian can be written in the following structure:
\begin{subequations}
\label{H_t}
\begin{align}
\mathcal{\hat{H}}& =\mathcal{\hat{H}}_{\text{M}}+\mathcal{\hat{H}}_{\text{AB}%
}+\mathcal{\hat{H}}_{\text{I}} \\
\mathcal{\hat{H}}_{\text{M}}& =-\hbar J\sum_{j=1}^{N-1}\left( \left\vert
j\right\rangle \left\langle j+1\right\vert +\text{h.c.}\right) -\mu
_{0}\left\vert N_{0}\right\rangle \left\langle N_{0}\right\vert \\
\mathcal{\hat{H}}_{\text{AB}}& =-\mu \left( \left\vert A\right\rangle
\left\langle A\right\vert +\left\vert B\right\rangle \left\langle
B\right\vert \right) \\
\mathcal{\hat{H}}_{\text{I}}& =-\hbar J_{A}\left( t\right) \left\vert
A\right\rangle \left\langle N_{0}-l\right\vert -\hbar J_{B}(t)\left\vert
B\right\rangle \left\langle N_{0}+l\right\vert +\text{h.c.},
\end{align}%
where $-J$ $(<0)$ is the coupling strength between nearest neighboring QDs
along the channel; $J_{A}\left( t\right) $ is the time dependent coupling
strength between $A$ and the $\left( N_{0}-l\right) $th site of the QD array
and is controlled by the sender, while $J_{B}(t)$ is the receiver controlled
coupling strength between $B$ and site $\left( N_{0}+l\right) $; $-\mu _{0}$
and $\mu $ are the on-site energy applied on the QDs; $\left\vert
j\right\rangle =c_{j}^{\dag }\left\vert 0\right\rangle $ represents the
Wannier state localized in the $j$-th quantum site for $j=A,1,2,...,N,B$.
For convenience, we consider the channel containing an odd number of QDs and
the gate voltage $-\mu _{0}$ is applied on the central dot, i.e. $%
N_{0}=(N+1)/2$. Note that equation (1) comprises three terms: the first
corresponds to the tight-binding chain with defect at $N_{0}$-site, the
second is the energy of QDs A, B, and the third term describes the
interaction Hamiltonian. In this paper we study the electron transfer from
QD-$A$ to QD-$B$ through the tight-binding array serving as quantum data
bus, and we denote the transfer distance $d=2l+3$. In this proposal the
propagation of the electron is driven by two time-dependent coupling
strengths, i.e., $J_{A}\left( t\right) $ and $J_{B}\left( t\right) $, which
are modulated in sinusoidal pulses

\end{subequations}
\begin{subequations}
\begin{align}
J_{A}\left( t\right) & =J_{0}\sin ^{2}\left( \frac{\pi t}{2t_{\max }}\right)
,  \label{J_A} \\
J_{B}\left( t\right) & =J_{0}\cos ^{2}\left( \frac{\pi t}{2t_{\max }}\right)
,  \label{J_B}
\end{align}
where $t_{\max }$ is the prescribed duration of QST; $J_{0}$ is the maximum
tunnelling rates between the two external QD and the defected chain. These
two pulses are illustrated in Fig. 2(a).

Firstly, let us perform a detail analysis of the peculiar properties of the
chain with single defect when used as a quantum channel for quantum state
transfer. Note that the Hamiltonian $\mathcal{\hat{H}}_{\text{M}}$ is
equivalent to a tight-binding problem with single diagonal impurity at site $%
N_{0}$. Let $\left\{ \left\vert \lambda _{n}\right\rangle \right\}$ and $%
\left\{ \lambda _{n}\right\} $ with $n=1,2,\ldots N-1$ be the sets of the
eigenstates of $\mathcal{\hat{H}}_{\text{M}}$, respectively. Then, $\mathcal{%
\hat{H}}_{\text{M}}$ can be rewritten to be

\end{subequations}
\begin{equation}
\mathcal{\hat{H}}_{\text{M}}=\sum_{n=0}^{N-1}\lambda _{n}\left\vert \lambda
_{n}\right\rangle \left\langle \lambda _{n}\right\vert .  \label{H_M}
\end{equation}

For $\mu _{0}=0$, the eigenstates are $\left\vert \lambda _{n}\right\rangle
=\sum_{j=1}^{N}\sin [(n+1)\pi j/(N+1)]\left\vert j\right\rangle $ with
energies $\lambda _{n}=2J\cos [(n+1)\pi /(N+1)]$, and where $n=0,1,2,...,N-1$%
. For non-zero $\mu _{0}$, we write the state in the single-particle Hilbert
space as $\left\vert \lambda _{n}\right\rangle
=\sum_{j=1}^{N}u_{n}(j)\left\vert j\right\rangle $. To see more precisely
what happens for $\mu _{0}\neq 0$, we solve the discrete-coordinate Schr\"{o}%
dinger equation

\begin{equation}
-J\left[ u_{n}(j-1)+u_{n}(j+1)\right] =\left( \mu _{0}\delta
_{j,N_{0}}+\lambda _{n}\right) u_{n}(j),  \label{E_eq}
\end{equation}
where $j\in \left[ 2,N-1\right] $. At the boundaries, we get slightly
different equations: $-Ju_{n}(2)=\lambda _{n}u_{n}(1)$ and $%
-Ju_{n}(N-1)=\lambda _{n}u_{n}(N)$.

It is well known that the $\delta$-type diagonal defect contributes exactly
one bound state of single particle case. To solve above equations, we assume
a usual solution by taking the mirror symmetry into consideration

\begin{equation}
u_{n}(j)\propto \left\{
\begin{array}{c}
\sin \left( k_{n}j\right) ,\text{ \ \ \ \ \ \ \ \ \ \ \ \ }j\leq N_{0} \\
(-1)^{n}\sin \left[ k_{n}(2N_{0}-j)\right] ,\text{ }j>N_{0}%
\end{array}
\right.  \label{U_n}
\end{equation}
where $k_{n}$ is the wave vector. By inserting this expression into
equations (\ref{E_eq}) and together with boundary condition, we obtain the
eigenvalues $\lambda _{n}=-2J\cos k_{n}$ in terms of wave vector $k_{n}$,
which obeys

\begin{equation}
\cot \left( k_{n}N_{0}\right) \sin k_{n}=\left\{
\begin{array}{c}
0,\text{ \ }n=\text{odd number} \\
\xi ,\text{ \ }n=\text{even number}%
\end{array}
\right. ,  \label{C_eq}
\end{equation}
where $\xi =\mu _{0}/2J$.

Based on the equation (\ref{C_eq}), one can get $N-1$ discrete real values $%
k_{n}$ in the interval $(0,\pi )$ and \textit{one} purely imaginary wave
vector. Together with the expression $\lambda _{n}=-2J\cos k_{n}$, the
eigenenergies corresponding to real wave vectors are included in the band $%
\left( -2J,2J\right) $. On the other hand, the purely imaginary wave vector
give rise to a out-of-band eigenenergy. Setting $k_{0}=iq$ and substituting
it into Eq. (\ref{C_eq}), we have $q=\ln \left[ \xi +\sqrt{\xi ^{2}+1}\right]
$. The wave vector $q$ yields the eigenvalue

\begin{equation}
\lambda _{0}=-2J\cosh q=-2J\sqrt{\xi ^{2}+1}.
\end{equation}
which splits off from the band. The corresponding localized state is given
by $\left\vert \lambda _{0}\right\rangle =\sum_{j=1}^{N}u_{0}(j)\left\vert
j\right\rangle $, with

\begin{equation}
u_{0}(j)=\Lambda ^{-1/2}e^{-q\left\vert N_{0}-j\right\vert }  \label{U_0}
\end{equation}
where $\Lambda =\cosh q/\sinh q$.

Until now, we have only discussed the solutions of Eq. (\ref{E_eq}) without
any external perturbation. In the thermodynamic limit $N_{0}\rightarrow
\infty $, the excited energies become a continuous energy band; it is not
hard to find that the energy gap between the ground state and the first
excited state (see Fig. 2(b)) is

\begin{equation}
\Delta =2J\left( \sqrt{\xi ^{2}+1}-1\right) .
\end{equation}

\begin{figure}[h!]
\center
\includegraphics[bb=86 398 354 779, width=7 cm, clip]{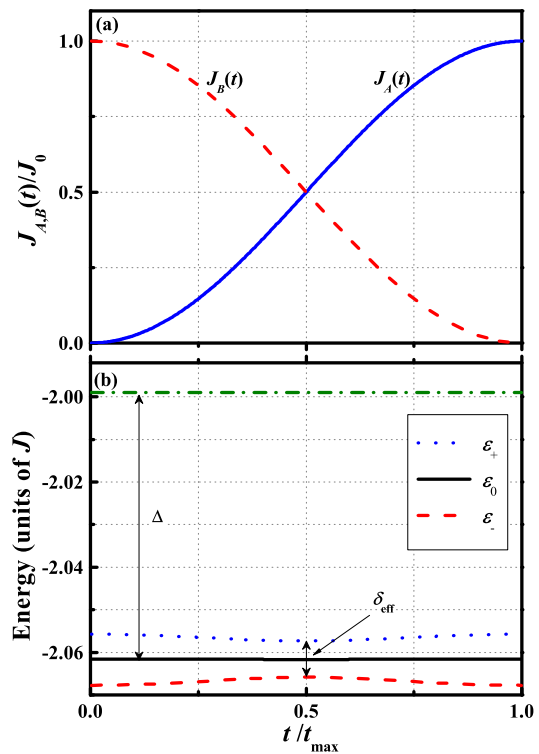}
\caption{(Color online) (a)The time-dependent tunnelling rates $J_{A}(t)$
and $J_{B}(t)$ as a function of time (in units of $J_{0}$), $J_{A}(t)$ is
the solid line and $J_{B}(t)$ is the dashed line. The pulse sequence is
applied in the counter-intuitive order. (b) The instantaneous eigenenergy
(in units of $J$) of the lowest four eigenstates of total Hamiltonian (1)
through the pulse shown in (a), which were obtained by direct numerical
diagonalization of the Hamiltonian. In the weak coupling limit, i.e. $%
J_{0}\ll J$ three lowest states is approximately equivalent to that a
triple-quantum-dot system.}
\end{figure}

\section{Adiabatic transport scheme}

\subsection{Effective Hamiltonian}

We now turn our attention to derive the effective Hamiltonian of the total
system when the couplings between QDs A, B (sender and receiver) and the
tight-binding QD array are weak. In the absence of the coupling between the
QDs A, B and the QD array ($J_{A}=J_{B}=0$) the ground states of the total
Hamiltonian (\ref{H_t}) are threefold degenerate for one electron problem by
setting $\mu =2J\sqrt{\xi ^{2}+1}$, i.e., the states $\left\vert
A\right\rangle $, $\left\vert \lambda _{0}\right\rangle $, and $\left\vert
B\right\rangle $ have the same energy $-\mu $. According to Eq. (\ref{J_A})
and (\ref{J_B}), The time-dependent tunnelling rates $J_{A}$ and $J_{B}$ are
varied in the interval $[0,J_{0}]$ as time process from $0$ to $t_{\max }$.
We assume the couplings between two side QDs and the bus are weak, i.e. $%
J_{0}\ll \Delta $, the Hamiltonian $\mathcal{\hat{H}}_{\text{I}}\left(
t\right) $ could be treated as perturbation within the time interval $%
[0,t_{\max }]$. Hence, the effective Hamiltonian can be derived by using
perturbation theory at any time during the process, which acts on the
subspace $\left[ \mathcal{G} \right] $ spanned by vectors $\left\vert
A\right\rangle $, $\left\vert \lambda _{0}\right\rangle $, and $\left\vert
B\right\rangle $.

In first-order degenerate perturbation theory, the matrix of the effective
Hamiltonian with states ordering $\left\{ \left\vert A\right\rangle
,\left\vert \lambda _{0}\right\rangle ,\left\vert B\right\rangle \right\} $
reads

\begin{equation}
\mathcal{\hat{H}}_{\text{eff}}=\left[
\begin{array}{ccc}
-\mu & -\Omega _{A}\left( t\right) & 0 \\
-\Omega _{A}\left( t\right) & -\mu & -\Omega _{B}\left( t\right) \\
0 & -\Omega _{B}\left( t\right) & -\mu%
\end{array}
\right]  \label{H_eff}
\end{equation}
where $\Omega _{\alpha }\left( t\right) =J_{\alpha }\left( t\right)
u_{0}(N_{0}-l)$, for $\alpha =A,B$. The eigenstates of the Hamiltonian of
the Eq.~(\ref{H_eff}) are

\begin{subequations}
\label{D_eff}
\begin{align}
\left\vert \mathcal{D}_{-}\left( t\right) \right\rangle & =\frac{1}{\sqrt{2}}
\left[ \sin \theta \left( t\right) \left\vert A\right\rangle -\left\vert
\lambda _{0}\right\rangle +\cos \theta \left( t\right) \left\vert
B\right\rangle \right] , \\
\left\vert \mathcal{D}_{0}\left( t\right) \right\rangle & =\cos \theta
\left( t\right) \left\vert A\right\rangle -\sin \theta \left( t\right)
\left\vert B\right\rangle , \\
\left\vert \mathcal{D}_{+}\left( t\right) \right\rangle & =\frac{1}{\sqrt{2}}
\left[ \sin \theta \left( t\right) \left\vert A\right\rangle +\left\vert
\lambda _{0}\right\rangle +\cos \theta \left( t\right) \left\vert
B\right\rangle \right] ,
\end{align}
where we have introduced the mixing angle $\theta \left( t\right) =\arctan %
\left[ J_{A}\left( t\right) /J_{B}\left( t\right) \right] $, and
corresponding energies are $\mathcal{E}_{0}=-\mu $, and $\mathcal{E}_{\pm
}=-\mu \mp \sqrt{ \left[ \Omega _{A}\left( t\right) \right] ^{2}+\left[
\Omega _{B}\left( t\right) \right] ^{2}}$.

In Fig. 2(b) we plot the four lowest eigenenergies of total Hamiltonian (1)
using the pulsing scheme given by Eq. (2) in weak coupling regime. To first
order in $J_{0}$, the perturbation Hamiltonian $\mathcal{\hat{H}}_{\text{I}}$
lifts degeneracy of the ground-state manifolds while the excited states $%
\left\vert \lambda _{n}\right\rangle $ are unaffected. It is worth
mentioning that the energy splitting $\delta _{\text{eff}}$ of the
ground-state subspace is proportional to $J_{0}$. This observation is
schematically shown in Fig. 2(b), in which $\Delta $ is the typical gap for
the unperturbed Hamiltonian $\mathcal{H}$ (i.e., $J_{0}=0$). In fact, the
weak coupling limit yields the inequality $\delta _{\text{eff}}\ll \Delta $
which will be equivalent condition for the purposes of perturbation.

One can see that, for weak coupling case, the total system reduces to and
effective three-level system that has a non-evolving "dark state", which can
serve as the vehicle for population transfer in a STIRAP-like procedure.
Note that the angle $\theta \left( t\right)$ is totally dependent on the
ratio of the two pulse strength and the pulse sequence used here are in the
counterintuitive ordering. It is well known that, for the counterintuitive
sequence of pulses, in which $J_{B}(t)$ precedes $J_{A}(t)$, one has $\theta
=0$ for $t=0$ and $\theta =\pi /2$ for $t=t_{\text{max}}$. With these
results, one can see that the state $\left\vert \mathcal{D}_{0}\left(
t\right) \right\rangle $ is $\left\vert A\right\rangle $ initially and goes
to $\left\vert B\right\rangle$ finally. The goal here is to study the
coherent quantum state transfer from state $\left\vert A\right\rangle $ to
state $\left\vert B\right\rangle $ by slowly varying the alternating pulse
sequence between QDs A, B and the chain to drive the state transfer. Here we
define the transfer distance $d$ to be the number of QDs between the two QDs
which the sender A and receiver B are respectively connected with, i.e., $%
d=2l+3$. The transfer fidelity depends on two aspects: (i) the validity of
the effective Hamiltonian (\ref{H_eff}) which is derived from perturbative
way and (ii) the dynamics follows the instantaneous eigenstate $\left\vert
\mathcal{D}_{0}\left( t\right) \right\rangle $ is whether or not adiabatic.
We now investigate these two aspects in the following, and $J_{0}$ and $\mu
_{0}$ are scaled by $J$ for simplicity.

\begin{figure}[tbp]
\center
\includegraphics[bb=73 264 388 737, width=7 cm, clip]{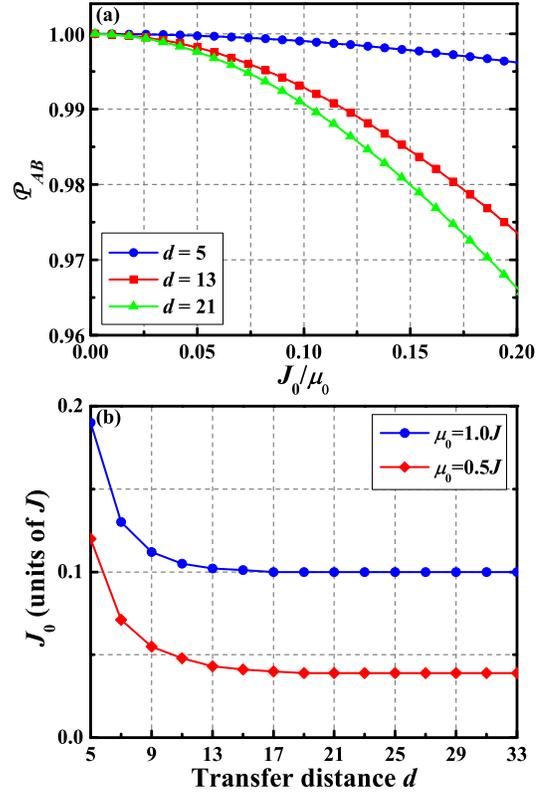}
\caption{(Color online) (a) The operator fidelity $\mathcal{P}_{AB}$ as a
function of $J_{0}/\protect\mu _{0}$ for $d=5$, $13$, and $21$. As the ratio
$J_{0}/\protect\mu _{0}$ increases the operator fidelity is decreased. (b)
The coupling $J_{0}$ as a function of $d$ for $\protect\mu _{0}=0.5J$, and $%
1.0J$ (bottom to top along $J_{0}$ axis) under the condition that the
operator fidelity greater than $99.5\%$. The other system parameter is $N=39$%
.}
\end{figure}

Bearing in mind the effective Hamiltonian (\ref{H_eff}) is the analytic
approximation of the total Hamiltonian (\ref{H_t}) and this approximation
holds when the energy splitting $\delta _{\text{eff}}$ caused by the $%
\mathcal{\hat{H}}_{\text{eff}}$ is smaller than the typical gap for the
unperturbed Hamiltonian $\mathcal{\hat{H}}$, i.e., $\delta _{\text{eff}}\ll
\Delta $. To investigate the range of validity about the above
approximation, we compare the instantaneous eigenstate $\left\vert \mathcal{D%
}_{0}\left( t\right) \right\rangle $ at time $t=t_{\max }/2$ of $\mathcal{%
\hat{H}}_{\text{eff}}$ with the density matrices reduced from the first
excited states of the total system. The density matrix corresponding to $%
\left\vert \mathcal{D}_{0}(t_{\max }/2)\right\rangle =\left( \left\vert
A\right\rangle -\left\vert B\right\rangle \right) /\sqrt{2}$ is

\end{subequations}
\begin{eqnarray}
\rho _{AB} &=&\left\vert \mathcal{D}_{0}(t_{\max }/2)\right\rangle
\left\langle \mathcal{D}_{0}(t_{\max }/2)\right\vert  \notag \\
&=&\frac{1}{2}\left[ \left\vert A\right\rangle \left\langle A\right\vert
+\left\vert B\right\rangle \left\langle B\right\vert -\left\vert
A\right\rangle \left\langle B\right\vert -\left\vert B\right\rangle
\left\langle A\right\vert \right] .
\end{eqnarray}
Moreover, we assign the state $\left\vert \Psi _{1}(t_{\max
}/2)\right\rangle $ denotes the instantaneous first-excited state for the
total Hamiltonian $\mathcal{\hat{H}}\left( t=t_{\max }/2\right)$. Then, the
operator fidelity is defined as

\begin{figure*}[tbp]
\center
\includegraphics[bb=95 578 355 782, width=5.5 cm, clip]{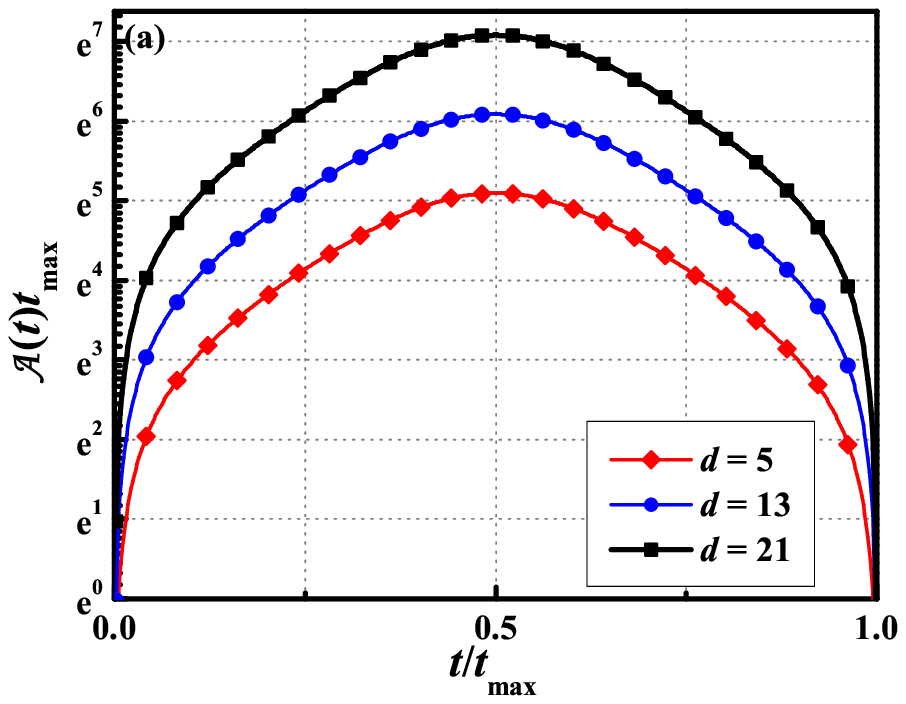} %
\includegraphics[bb=95 578 353 782, width=5.5 cm, clip]{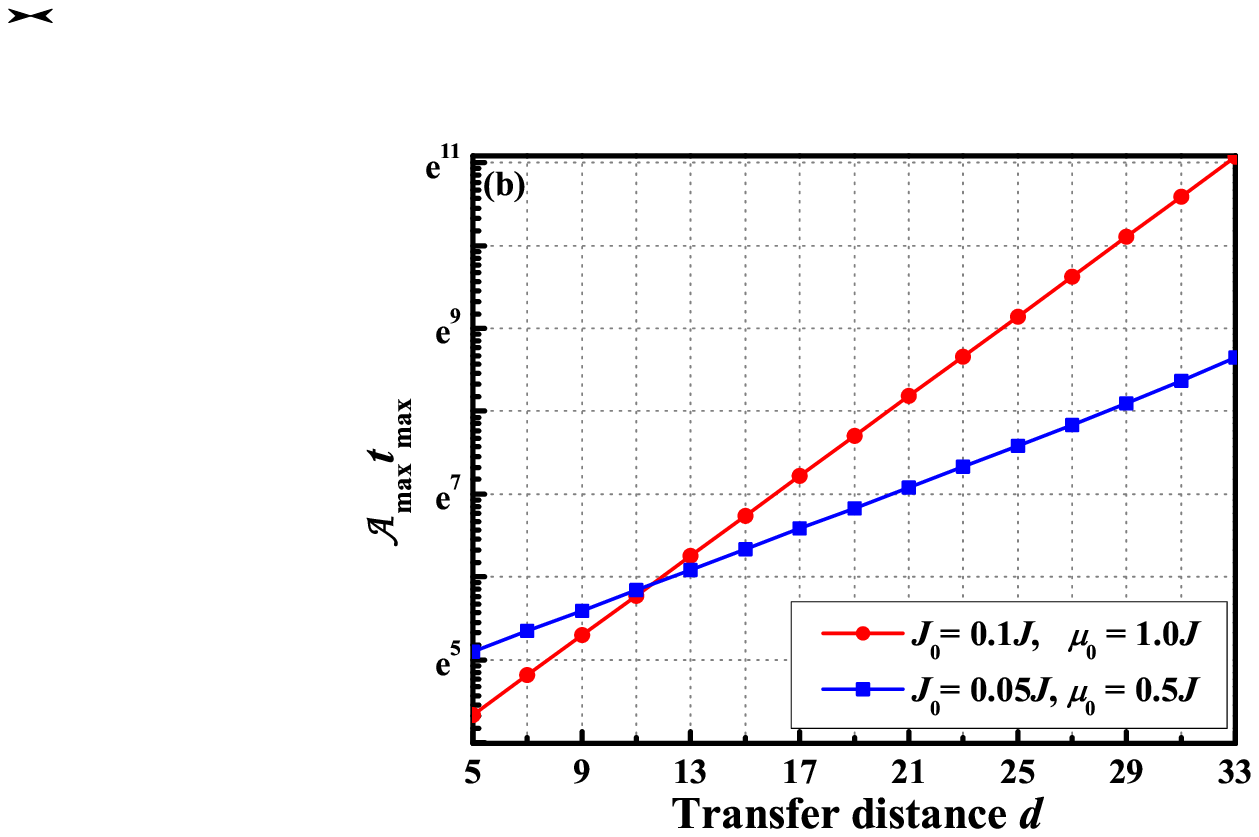} %
\includegraphics[bb=95 578 355 782, width=5.5 cm, clip]{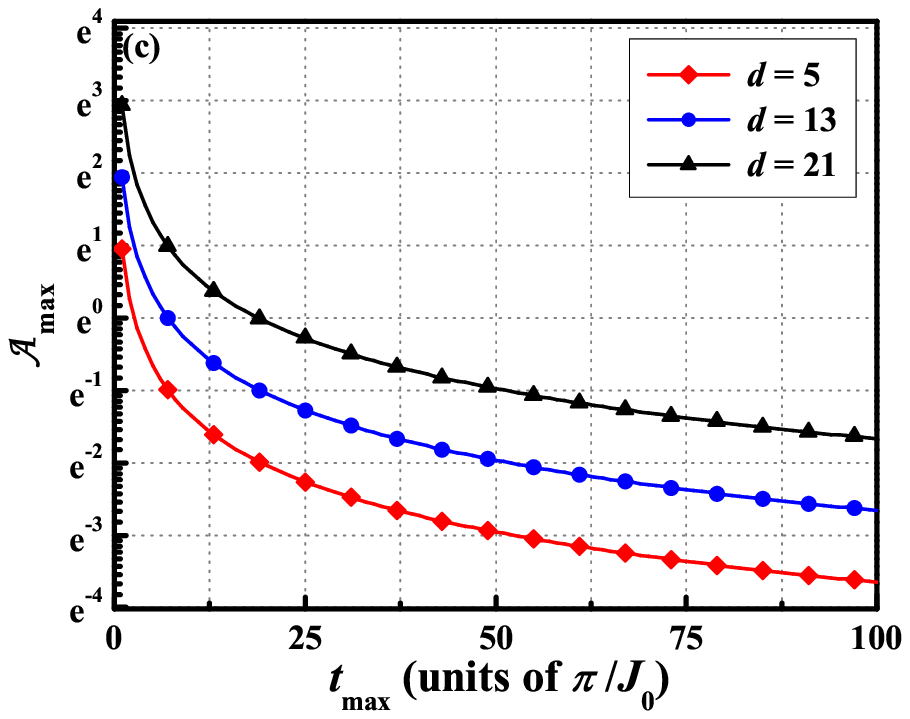}
\caption{(Color online) (a) Adiabaticity $\mathcal{A}\left( t\right) t_{\max
}$ as a function of time in the time interval $t\in \lbrack 0,t_{\max }]$
corresponding to the pulse shapes as in Eq. (2). The results show that the
adiabaticity parameter $\mathcal{A}\left( t\right) t_{\max }$ is largest at
the crossing point of the two pulses. The parameters we chosen are $N=39$, $%
J_{0}=0.05J$, and $\protect\mu _{0}=0.5J$. (b) Maximum adiabaticity $%
\mathcal{A}_{\max }t_{\max }$ through the protocol as a function of transfer
distance $d$ for $N=39$. The parameters is $J_{0}=0.05J$, $\protect\mu%
_{0}=0.5J$ (squares) and $J_{0}=0.1J$, $\protect\mu _{0}=1.0J$ (circles). As
the transfer distance increases, the adiabaticity parameter increases and
the grow of adiabaticity of the large $\protect\mu _{0}$ is faster than the
small one. (c) The maximum adiabaticity parameter $\mathcal{A}_{\max }$ as a
function of $t_{\max }$. As $t_{\max }$ is increased $\mathcal{A}_{\max }$
is decreased, indicating that better fidelity transfer can be achieved for
longer total transfer time.}
\label{fig4}
\end{figure*}

\begin{equation}
\mathcal{P}_{AB}=\left( \text{Tr}\sqrt{\rho _{AB}^{1/2}\rho _{R}\rho
_{AB}^{1/2}}\right) ^{2},
\end{equation}
where $\rho _{R}=$Tr$_{M}\left( \left\vert \Psi _{1}(t_{\max
}/2)\right\rangle \left\langle \Psi _{1}(t_{\max }/2)\right\vert \right) $,
and Tr$_{M}$ means the trace over the variables of the tight-binding array. $%
\mathcal{P}_{AB}$\ is sensitive to two parameters, i.e., the ratio of $%
J_{0}/\mu _{0}$ and the transfer distance $d$. In the following discussions,
we will investigate how the above external parameters influence the operator
fidelity of the dark state. Fig. 3(a) shows the dependence of $\mathcal{P}%
_{AB}$ on both $J_{0}/\mu _{0}$ and $d$ for the system with $N=39$. We can
see that the operator fidelity improves\ by decreasing $J_{0}/\mu _{0}$.
Moreover, we note that, increasing the value of $d$ there is a slight shift
of $\mathcal{P}_{AB}$ for a given $J_{0}/\mu _{0}$, which means that as the
transfer distance increases one need to decrease the ratio $J_{0}/\mu _{0}$
if we want high operator fidelity to hold. To obtain high quality of
operator fidelity ($\mathcal{P}_{AB}\geq 99.5$), we compute the ratio $%
J_{0}/\mu _{0}$ versus transfer distance $d$ with two different impurity
on-site energy $\mu _{0}$; these results are shown in Fig. 3(b). In this
figure, one can see that taking $J_{0}/\mu _{0}\leq 0.1$, the effective
Hamiltonian agrees very well with the exact solution obtained from numerical
calculations. Therefore, we have shown the weak coupling effective
Hamiltonian (\ref{H_eff}) is a very good approximation of the exact model.

\subsection{Adiabatic condition}

To realize high-fidelity QST, we require that the system remains in its dark
state $\left\vert \mathcal{D}_{0}\left( t\right) \right\rangle $, without
loss of population from this state to the neighboring states, i.e., $%
\left\vert \mathcal{D}_{\pm }\left( t\right) \right\rangle $. The
adiabaticity parameter defined for this scheme is

\begin{eqnarray}
\mathcal{A}\left( t\right) &=&\frac{\left\vert \left\langle \mathcal{D}
_{+}\left( t\right) \right\vert \partial _{t}\mathcal{\hat{H}}_{\text{eff}
}\left\vert \mathcal{D}_{0}\left( t\right) \right\rangle \right\vert } {%
\left\vert \mathcal{E}_{+}-\mathcal{E}_{0}\right\vert ^{2}}  \notag \\
&=&\frac{\left\vert \dot{\Omega}_{A}\left( t\right) \Omega _{B}\left(
t\right) -\dot{\Omega}_{B}\left( t\right) \Omega _{A}\left( t\right)
\right\vert }{\sqrt{2}\left[ \Omega _{A}^{2}\left( t\right) +\Omega
_{B}^{2}\left( t\right) \right] ^{3/2}}.
\end{eqnarray}
The time dependence of the parameter $\mathcal{A}\left( t\right) t_{\max }$
is illustrated in Fig. 4(a). Obviously the adiabaticity parameter $\mathcal{A%
}\left( t\right) t_{\max }$ reaches maximum at the crossing point of the two
pulses. At this point one has $\Omega _{A}\left( t_{\max }/2\right) =\Omega
_{B}\left( t_{\max }/2\right) =J_{0}u_{0}(N_{0}-l)/2$\ and $\dot{\Omega}%
_{A}\left( t_{\max }/2\right) =-\dot{\Omega}_{B}\left( t_{\max }/2\right)
=\pi J_{0}u_{0}(N_{0}-l)/2t_{\max } $. By using the form of the pulses as
given in Eq. (2), the above equation gives rise to the simple form
\begin{equation}
\mathcal{A}_{\max }=\frac{\pi }{J_{0}u_{0}(N_{0}-l)t_{\max }}.  \label{A_max}
\end{equation}
The analytical expression for maximum adiabaticity is helpful for estimating
the quantum state transfer time $t_{\max }$. For adiabatic evolution of the
system we require $\mathcal{A}_{\max }\ll 1$. According to Eq. (\ref{A_max}%
), the total transfer time should satisfy $t_{\max }\gg \pi /\left[
J_{0}u_{0}(N_{0}-l)\right] $. In Fig. 4(b) we present the dependence of the $%
\mathcal{A}_{\max }t_{\max }$ on the transfer distance $d$ for a system with
$N=39$. Obviously one sees the increase in the adiabaticity parameter with
increasing $d$. Moreover, one can see that the bigger is the $\mu _{0}$, the
faster is the growth of $\mathcal{A}_{\max }t_{\max }$. The reason is that
by increasing $\mu _{0}$, the localization effect of Eq.~(\ref{U_0}) is
enhanced. Figure 4(b) also reflects the fact that the optimal transfer time
needs to be increased with increasing $d$. The smaller the $\mu _{0}$ is,
the slower the growth rate will be. Furthermore, the dependence of the
adiabaticity parameter on the total protocol time $t_{\max }$ is plotted in
Fig. 4(c). With longer $t_{\max }$, and hence lower $\mathcal{A}_{\max }$,
the transported electron is more likely to remain in the desired $\left\vert
\mathcal{D}_{0}\left( t\right) \right\rangle $ state resulting in better
fidelity transfer.

\subsection{Coherent state transfer}

\begin{figure*}[tbp]
\center
\includegraphics[bb=95 199 356 779, width=7 cm, clip]{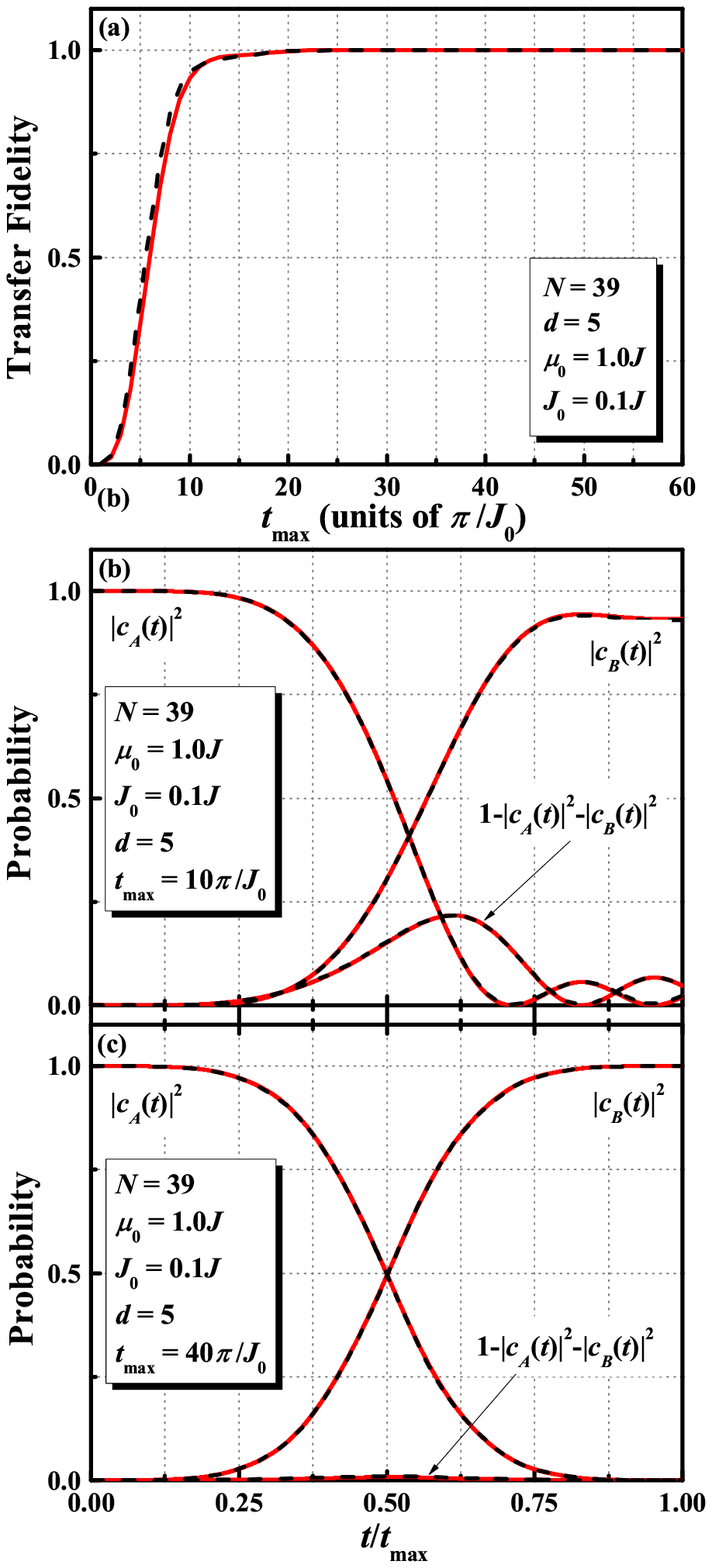} %
\includegraphics[bb=95 199 356 779, width=7 cm, clip]{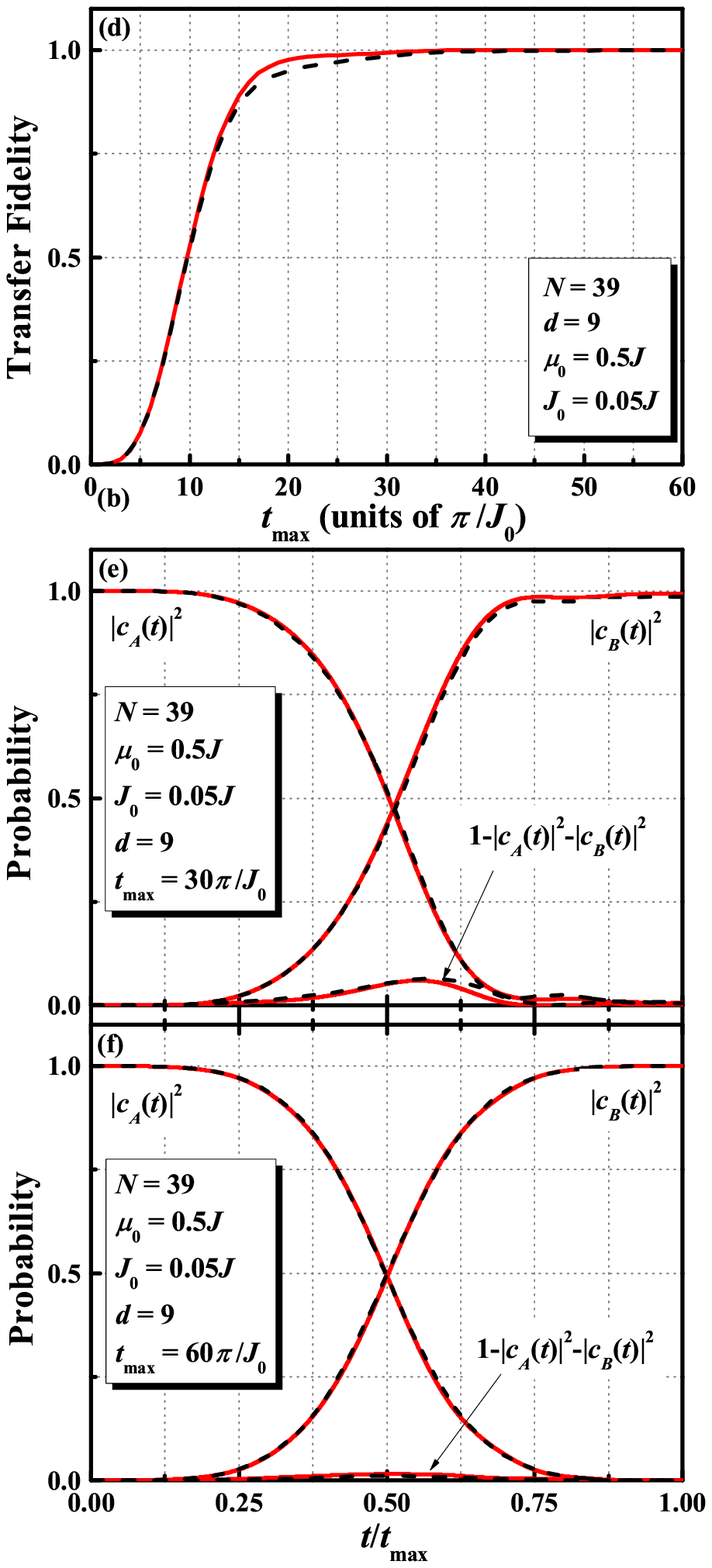}
\caption{(Color online) (a) The transfer fidelity $F$ as a
function of $t_{\max }$ (in units of $\protect\pi /J_{0}$) for $d=5$. The
parameters we take is $N=39$, $\protect\mu _{0}=1.0J$, and $J_{0}=0.1J$. The
red solid curves correspond to the approximate results via perturbation
theory and black dotted curves are the exact numerical results for the
complete Hilbert space. (b) The time evolution of the probabilities induced
by the pulses in Fig. 2(a) for $d=5$ and $t_{\max }=10\protect\pi /J_{0}$.
(c) is the same as (b) but for $t_{\max }=40\protect\pi /J_{0}$. (d) The
same as in (a), but for $d=9$, $\protect\mu _{0}=0.5J$, and $J_{0}=0.05J$.
(e)-(f) is same as (b)-(c) but for $d=9$. To get high fidelity transfer,
more time is required for longer transfer distance.}
\label{fig5}
\end{figure*}

To simulate the analog of STIRAP protocol we initialize the device so that
the particle occupies site $\left\vert A\right\rangle $ at $t=0$, i.e., the
total initial state is $\left\vert \Psi \left( 0\right) \right\rangle
=\left\vert A\right\rangle $, and apply the alternating pulse sequence [see
Eq. (2)] in the counterintuitive order. The evolution of the wave function
is described by Schr\"{o}dinger equation
\begin{equation}
i\frac{\partial }{\partial t}\left\vert \Psi \left( t\right) \right\rangle =%
\mathcal{\hat{H}}_{\text{eff}}\left\vert \Psi \left( t\right) \right\rangle ,
\label{S_eq}
\end{equation}%
which creates a coherent superposition: $\left\vert \Psi \left( t\right)
\right\rangle =c_{A}(t)\left\vert A\right\rangle +c_{0}(t)\left\vert \lambda
_{0}\right\rangle +c_{B}(t)\left\vert B\right\rangle $. Substituting the
superposition form of $\left\vert \Psi \left( t\right) \right\rangle $ into
the Schr\"{o}dinger equation, we get equations of motion for the probability
amplitudes
\begin{eqnarray}
i\dot{c}_{A}(t) &=&-\Omega _{A}\left( t\right) c_{0}(t),  \notag \\
i\dot{c}_{0}(t) &=&-\Omega _{A}\left( t\right) c_{A}(t)-\Omega _{B}\left(
t\right) c_{B}(t), \\
i\dot{c}_{B}(t) &=&-\Omega _{B}\left( t\right) c_{0}(t),  \notag
\end{eqnarray}%
where the dot denotes the time derivative.

A measure of the quality of this protocol after the pulse sequences is given
by the transfer fidelity
\begin{equation}
F=\left\vert \left\langle B\right. \left\vert \Psi \left( t_{\max }\right)
\right\rangle \right\vert ^{2}=\left\vert c_{B}(t_{\max })\right\vert ^{2}.
\end{equation}%
To investigate QST between QDs $A$ and $B$, we numerically solve the
time-dependent Schr\"{o}dinger equation for the multi-dot system with $N=39$%
. Here we compare the results obtained from two alternative approaches. In
the first case, we numerically integrated Eq. (\ref{S_eq}) with the initial
condition $\left\vert \Psi \left( 0\right) \right\rangle =\left\vert
A\right\rangle $. In the second case, we adopt total Hamiltonian (1) to
replace the effective Hamiltonian and we perform a numerical simulation
using the total Hamiltonian. In this case the computation takes place in
full Hilbert space with basis $\left\{ \left\vert A\right\rangle ,\left\vert
1\right\rangle ,\left\vert 2\right\rangle ,\ldots ,\left\vert N\right\rangle
,\left\vert B\right\rangle \right\} $. Fig. 5(a) shows transfer fidelity $F$
as a function of $t_{\max }$ for $d=5$, $\mu _{0}=1.0J\,$\ and $J_{0}=0.1J$.
If we choose $t_{\max }\geq 19\pi /J_{0}$, the transfer fidelity $F$\ will
be larger than $99.5\%$. To illustrate the process of QST, we exhibit in
Fig. 5(b) and 5(c) the time evolution of the probabilities. We get perfect
state transfer if we choose the transfer time longer enough and the
populations on the QD \textit{A} and QD \textit{B} are exchanged in the
expected adiabatic manner. For the case $d=9$ with $\mu _{0}=0.5J\,$\ and $%
J_{0}=0.05J$, it is shown in Fig. 5(d)-(f) that to ensure $F\geq 99.5\%$ the
optimal transfer time is about $31\pi /J_{0}$. For comparison, we also plot
in Fig. 5 the result of the exact numerical results (dashed curves) for the
full Hilbert space calculation. Obviously, our three-state effective
Hamiltonian describes the quantum state evolution very well.

In order to provide the most economical choice of total transfer time $%
t_{\max }$ for reaching high transfer efficiency, we perform numerical
analysis of the relation between $t_{\max }$ and $d$. For a given tolerable
transfer error $1-F=0.5$, we plot in Fig. 6 the minimum time varies as a
function of transfer distance $d$. Clearly, the time required for
near-perfect transfer depends on $d$ in an exponential fashion. Intuitively,
this behavior results from the fact that the ground state of medium
Hamiltonian $\mathcal{\hat{H}}_{\text{M}}$ is exponentially localized due to
the existence of defect $\mu _{0}$, leading to the exponential decrease of
the energy splitting $\delta _{\text{eff}}$ as $d$ increases; and this
effect is stronger for larger $\mu _{0}$ and weaker for smaller $\mu _{0}$.
In these sense, the negative effects of $t_{\max }$ on the transfer distance
can be partially compensated by reducing the defect energy $\mu _{0}$. It is
conceivable that $t_{\max }$ will scale linearly with $d$ in the limit $\mu
_{0}\rightarrow 0$.

\begin{figure}[tbp]
\center
\includegraphics[bb=85 470 384 708, width=7 cm, clip]{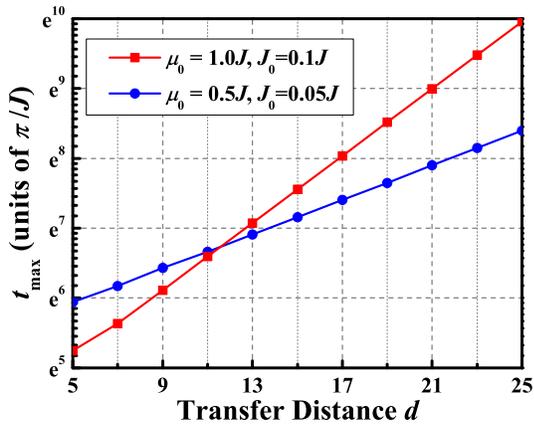}
\caption{(Color online) The plot of distance dependence of transfer time $%
t_{\max }$ for a chosen tolerable transfer error $1-F=0.5\%$. The lines are
only guides to the eyes. Notice the exponential increase of $t_{\max }$ as a
function of the distance. In order to make results comparable, all times are
scaled in units of $\protect\pi /J$.}
\label{fig6}
\end{figure}

\subsection{Robustness of state transfer}

We have shown that under appropriate system parameters, the total
Hamiltonian $\mathcal{\hat{H}}$ can be mapped to a three-level effective
Hamiltonian $\mathcal{\hat{H}}_{\text{eff}}$, which establish an effective
STIRAP pathway for realizing long-range QST. Not only efficiency but also
robustness against technical and fundamental noise is important for a scheme
to be applicable in quantum-information processing and quantum computing.
There are two central concerns for the QST protocol in experiments,
decoherence and imperfect experimental implementations.

In this section, we are going to consider an realistic model which is more
close to experimental implementation and analyze the robustness of this
scheme to unavoidable imperfections. In particular, we assume that the
Hamiltonian has a random but constant offset fluctuation in the couplings,
i.e., replacing the couplings in Eq. (1b) with $J\rightarrow J\left(
1+\delta \epsilon _{j}\right) $. The total Hamiltonian is therefore
\begin{widetext}
\begin{eqnarray}
\mathcal{\hat{H}}^{\prime } &=&\sum_{j=1}^{N-1}-J\left( 1+\delta \epsilon
_{j}\right) \left\vert j\right\rangle \left\langle j+1\right\vert -\frac{\mu
_{0}}{2}\left\vert N_{0}\right\rangle \left\langle N_{0}\right\vert -\frac{\mu }
{2}\left( \left\vert A\right\rangle \left\langle A\right\vert
+\left\vert B\right\rangle \left\langle B\right\vert \right)   \notag \\
&&-J_{A}\left( t\right) \left\vert A\right\rangle \left\langle
N_{0}-l\right\vert -J_{B}(t)\left\vert B\right\rangle \left\langle
N_{0}+l\right\vert +\text{h.c.},
\end{eqnarray}
\end{widetext}where $\delta $ is the maximum coupling offset bias relative
to $J$; $\epsilon _{j}$ is drawn from the standard uniform distribution in
the interval $[-1,1]$ and all $\epsilon _{j}$ are completely uncorrelated
with all sites along the medium chain.

Due to the existence of imperfections, the former effective Hamiltonian (\ref%
{H_eff}) is no longer hold. We now consider the two effects together with
the master equation \cite{ADG04}%
\begin{equation}
\dot{\rho}\left( t\right) =-i\left[ \mathcal{\hat{H}}^{\prime }\left(
t\right) ,\rho \left( t\right) \right] -\Gamma \left\{ \rho \left( t\right) -%
\text{diag}\left[ \rho \left( t\right) \right] \right\} ,  \label{MEQ}
\end{equation}%
where $\Gamma $ is the pure dephasing reate. To determine the robustness of
the perturbed situation, we numerically integrate above equation with the
electron initialized in QD-\textit{A} to be transported: $\rho \left(
t=0\right) =\left\vert A\right\rangle \left\langle A\right\vert $. At the
end of the computation $\left( t=t_{\max }\right) $, we obtain the density
matrix $\rho \left( t_{\max }\right) $. The problem in the following we
concern will be to evaluate the transfer fielity
\begin{equation}
\mathcal{F}=\text{Tr}\left[ \rho \left( t_{\max }\right) \rho _{B}\right]
\end{equation}%
where $\rho _{B}=\left\vert B\right\rangle \left\langle B\right\vert $.

Setting $\mu _{0}=1.0J$, $J_{0}=0.1J$, and $d=5$, figure 7(a) shows the
solutions of the master equation (\ref{MEQ}) for $\Gamma =0$ and for
different values of maximum coupling offset $\delta $, in which we have
chosen to report the transfer fidelity $\mathcal{F}$ as a function of the
total duration time $t_{\max }$ of QST. Bearing in mind that a total
duration time $t_{\max }$ greater than $19\pi /J_{0}$ can guarantee the
perfect state transfer for ideal case $\left( \text{i.e.},\Gamma =0\text{
and }\delta =0\right) $. As expect, this approach is robustly insensitive to
weak fluctuations $\left( \delta \leq 0.1\right) $ of the couplings. By
increasing $\delta $, the negative effects on transfer fidelity become more
and more pronounced, and this negative influence can be compensated when the
duration of the process is chosen to be long enough. This means that the
scheme allows one to increase the transfer fidelity arbitrarily close to
unity, without the need for a precise control of the couplings. In Fig.
7(b), we show the effects of dephasing on transfer fidelity. When dephasing
is considered, perfect QST cannot be achieved. For optimum value of $t_{\max
}$, transfer fidelity has a maximum value $\mathcal{F}_{\max }$ which
decreases when $\Gamma $ is increased. For $\Gamma =0.001J_{0}$, the
optimum value for $t_{\max }$ is $t_{\max }=21\pi /J_{0}$ with $\mathcal{F}%
_{\max }=0.99$. When $\Gamma =0.005J_{0}$, $\mathcal{F}_{\max }$ reaches $%
0.95$ at $t_{\max }=18\pi /J_{0}$. The optimum value of $t_{\max }$\ is
slightly shorter than ideal case because dephasing will have more time to
destroy the coherent transfer as $t_{\max }$\ increases.
\begin{figure}[tbp]
\center
\includegraphics[bb=73 249 387 697, width=7 cm, clip]{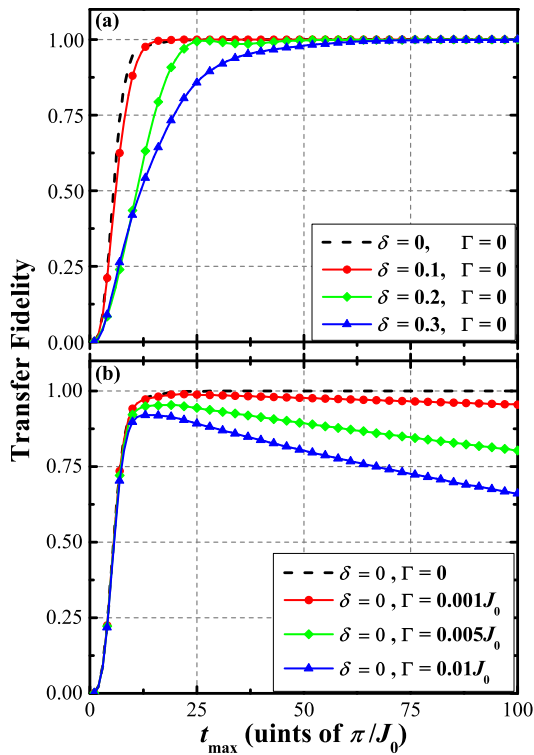}
\caption{(Color online) Transfer fidelity $\mathcal{F}$ as a function of total transfer
time $t_{\max }$ for (a) $\Gamma =0$ and different values of $\delta $; (b) $\delta =0$ and different values of $\Gamma$.
The parameter values we take is $N=39$, $\protect\mu _{0}=1.0J$, $J_{0}=0.1J$, and $d=5$.}
\label{fig7}
\end{figure}

\section{Summary}

The central idea of this work is to model a solid-state adiabatic quantum
communication protocol suitable for high fidelity robust multiple-range QST
and quantum information swapping. It has been shown that high fidelity
two-way QST can be realized at various different distances by introducing an
\emph{N}-site tight-binding QD array as quantum data bus. We first
demonstrate that the tight-binding chain with a diagonal defect has a
non-vanishing energy gap above the ground state in the single-particle
subspace; and this defect produces an eigenstate exponentially localized at
the defect. Our approach to realize high-fidelity QST is based on the fact
that the two information exchange QDs are resonantly coupled to the zeroth
eigen-mode of the quantum data bus. By treating the weak coupling as
perturbation, the system can be reduced to a three-level system by the
first-order terms in the perturbative expansion, which enables us to perform
an effective three-level STIRAP. Then we present that it is possible to
transfer an arbitrary quantum state driven by adiabatically modulating two
side couplings. For proper choices of the system parameters, perfect
adiabatic QST can be obtained, which has been confirmed by exact numerical
simulations. Moreover, for an increasing transfer distance, we find that the
evolution time displays an exponential dependence on the transfer distance.
However, this negative effect can be suppressed by reducing the defect
energy $\mu _{0}$. Finally, the robustness of the scheme to fabrication
disorder and dephasing is numerically demonstrated.

Comparing to the existing long-rang QST schemes, our proposal has the
following advantages: i) the requirement of tunnelling control are
minimized. This means that our scheme provides an efficient alternative
long-range QST scheme without performing many tunnelling operations. ii)
different transfer distance can be achieved by changing the connecting site
of two side QDs, and no additional QDs are needed. Our proposal provides a
novel scheme to implement ordinary STIRAP protocols in many-body solid-state
systems in the realization of high-fidelity multiple-range QST.

\section*{Acknowledgements}

This work is supported by the National Natural Science Foundation of China
(Grant Nos. 11105086, 11204162), and the SDUST Research Funds (2012KYTD103
and 2014JQJH104), as well as the National Science Foundation through awards
PHY-1505189 and INSPIRE PHY-1539859.

\end{document}